\documentclass[conference]{IEEEtran}
\IEEEoverridecommandlockouts
\usepackage{cite}
\usepackage{amsmath,amssymb,amsfonts}
\usepackage{algorithmic}
\usepackage{graphicx}
\usepackage{textcomp}
\usepackage{multirow}
\usepackage[table,xcdraw]{xcolor}
\def\BibTeX{{\rm B\kern-.05em{\sc i\kern-.025em b}\kern-.08em
    T\kern-.1667em\lower.7ex\hbox{E}\kern-.125emX}}
    
\begin{document}

\title{A CNN-Transformer Deep Learning Model for Real-time Sleep Stage Classification in an Energy-Constrained Wireless Device
}

\author{
\IEEEauthorblockN{Zongyan Yao}
\IEEEauthorblockA{
\textit{University of Toronto}\\
Toronto, Canada \\
zongyan.yao@mail.utoronto.ca}
\and
\IEEEauthorblockN{Xilin Liu}
\IEEEauthorblockA{
\textit{University of Toronto}\\
Toronto, Canada \\
xilinliu@ece.utoronto.ca}
}

\maketitle

\begin{abstract}

This paper proposes a deep learning (DL) model for automatic sleep stage classification based on single-channel EEG data. The DL model features a convolutional neural network (CNN) and transformers. The model was designed to run on energy and memory-constrained devices for real-time operation with local processing. The Fpz-Cz EEG signals from a publicly available Sleep-EDF dataset are used to train and test the model. Four convolutional filter layers were used to extract features and reduce the data dimension. Then, transformers were utilized to learn the time-variant features of the data. To improve performance, we also implemented a subject specific training before the inference (i.e., prediction) stage. With the subject specific training, the F1 score was 0.91, 0.37, 0.84, 0.877, and 0.73 for wake, N1- N3, and rapid eye movement (REM) stages, respectively. The performance of the model was comparable to the state-of-the-art works with significantly greater computational costs. We tested a reduced-sized version of the proposed model on a low-cost Arduino Nano 33 BLE board and it was fully functional and accurate. In the future, a fully integrated wireless EEG sensor with edge DL will be developed for sleep research in pre-clinical and clinical experiments, such as real-time sleep modulation.

\end{abstract}

\section{Introduction}
\par Sleep quality and health are closely related; therefore, it is important to understand one’s sleep quality to improve health condition. A measurement of sleep quality is the time spent in each sleep stage. There are five sleep stages, which are wake, N1, N2, N3, and REM, with each stage progressively deeper sleep. Most of the sleep occurs between stages N1 and N3 \cite{11}. The clinical evaluation of sleep stages is performed by polysomnogram (PSG), a procedure that records one’s electroencephalogram (EEG), electrooculogram (ECG), and other physiological features. Medical professionals will manually classify their sleep stages over time according to one or more of the features mentioned above. 
\begin{figure}[!ht]
    \centering
    \includegraphics[width=1\linewidth]{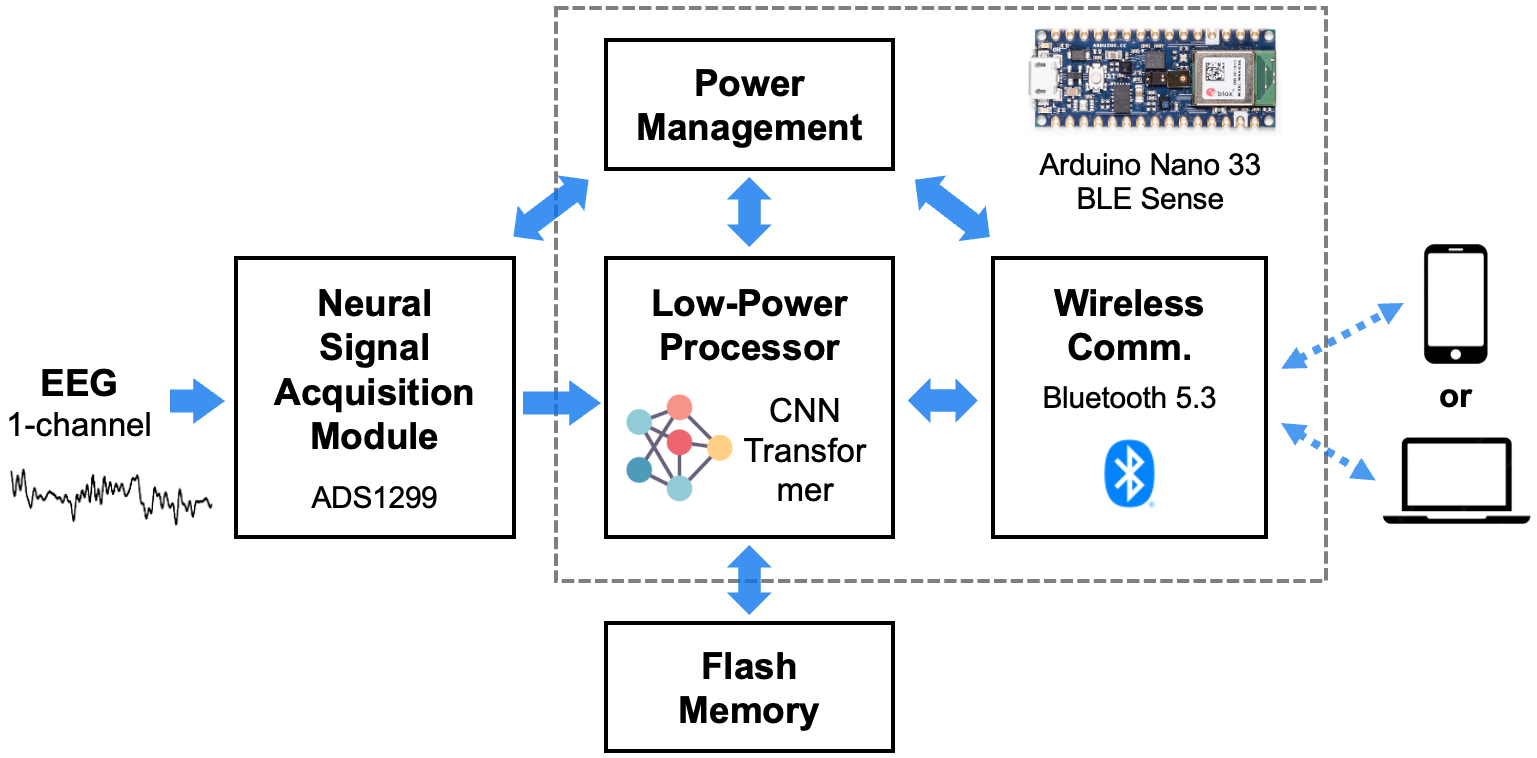}
    \caption{An overview of the envisioned wireless device for real-time sleep stage classification using edge DL. This paper focuses on the development and deployment of the DL model. }
    \label{fig:interface}
\end{figure}
\par With the development of electronic technology and machine intelligence, wearable devices, such as smartwatches, can measure user biosignals and potentially classify their sleep stages. However, the cost of these devices is high and the performance of sleep stage classification is limited. In addition, classification often requires the transmission of data to mobile phones or the cloud, raising concerns about cybersecurity \cite{liu2021energy,liu2020fully}. High-quality real-time sleep classification and sleep modulation still needs to be performed in sleep laboratories. There is a need for low-cost at-home sleep monitoring devices that can perform sleep stage classification on device, and potentially use the sleep stage to generate auditory stimulation for treating sleep disorders or enhancing sleep quality \cite{liu2021system}. 


\par In this paper, we develop a lightweight DL model for running on devices with restricted energy and memory, such as microcontrollers \cite{liu2014pennbmbi}. There are two main constraints to the development of the model for hardware. The first constraint is that the size of the model will be limited by the memory resources available on the hardware, including the non-volatile Flash memory for model storage and the random-access memory (RAM) for model computing \cite{liu2021edge}. The second constraint is that the computational demand of the model will be limited by the clock rate, bit width, and computational capabilities (such as floating point or fixed point) of the device \cite{liu2016virtual}. Key trade-offs are between model performance and complexity. In this work, we developed a DL model that can run on a low-power wireless microcontroller, based on a low-cost Arduino Nano 33 BLE development board. Despite its small size, our model achieved performances comparable to the state-of-the-art during a validation using a publicly available sleep dataset.

\par Fig.~\ref{fig:interface} shows the overall block diagram of the envisioned fully integrated sleep stage classification device featuring the developed DL model. The device will be miniature in size and fully self-contained. It can enable a wide range of sleep research in pre-clinical and clinical studies.

\section{Methods}

\begin{figure*}
    \centering
    \includegraphics[width=.7\linewidth]{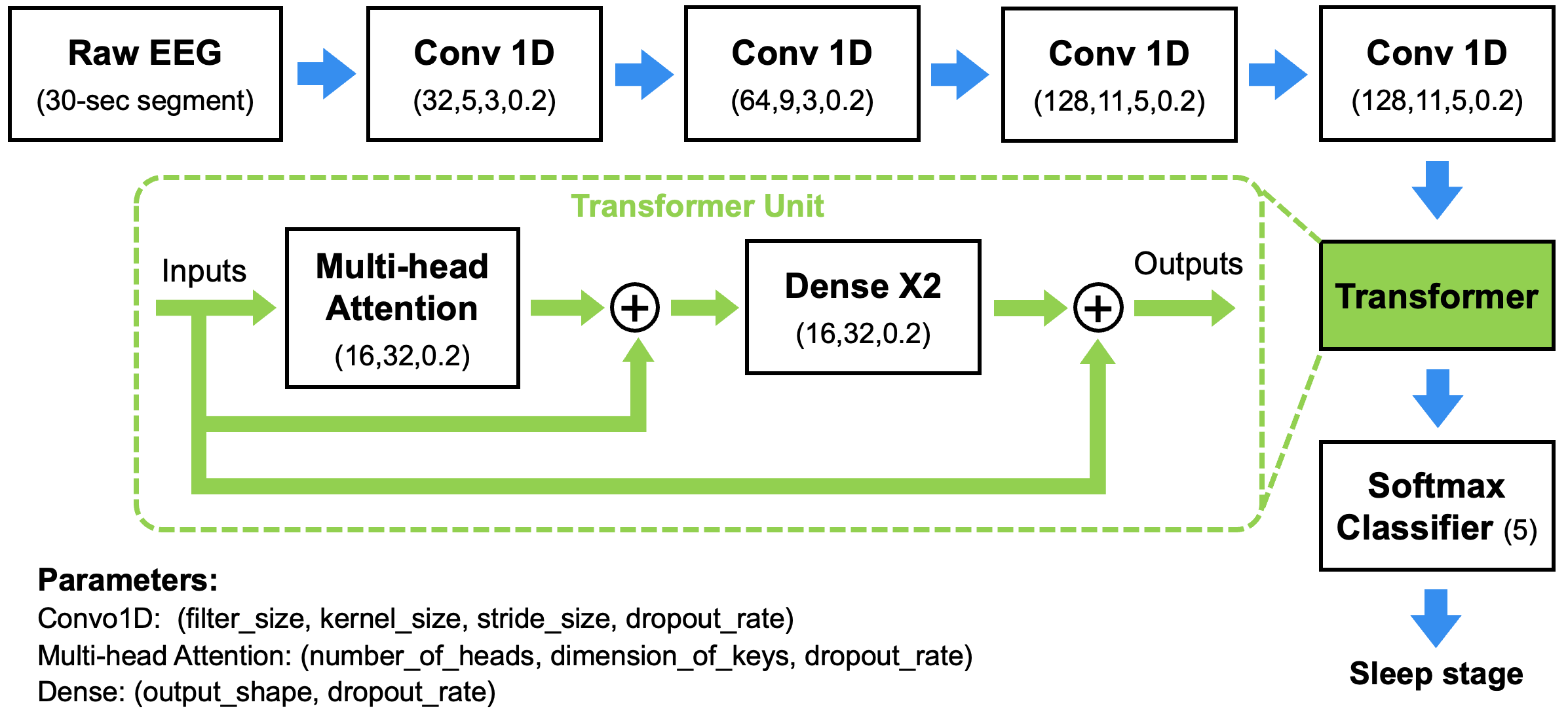}
    \caption{Architecture of the proposed CNN-Transformer DL model for real-time sleep stage classification using single channel EEG signal. }
    \label{fig:model}
\end{figure*}

\subsection{Dataset}
\par Sleep-EDF Expanded Database (version 1, published in 2013) contains 197 whole-night Polysomnographic (PSG) sleep recordings \cite{1,2}. It contains two subsets, the Sleep Cassette Study (SC) and Sleep Telemetry Study (ST). Data from the Sleep Telemetry Study was obtained in 1994 to study the effects of temazepam on sleep. Since we are proposing a model to classify the stages of sleep of healthy people, we will not use data from this subset. 
Our experiments will be performed on the SC subset.

\par Two 20-hour PSG recordings were taken for 77 subjects between the age of 25 and 101. The first nights of subjects 36 and 52, and the second night of subject 13 were lost. The PSG recordings contain three channels of EEG signals, one channel of EOG and chin EMG signals, oronasal airflow, rectal body temperature, and event marker. To reduce our model’s size, we only use Fpz-Cz EEG signals as input to our model. The EEG signals were sampled at 100 Hz. Each of the 30-second segments of the signals was labeled by well-trained sleep experts. There are eight stages, N1, N2, N3, N4, Wake, REM, MOVEMENT, UNKNOWN). To make our results consistent and comparable with previous studies \cite{3,4,5}, we preprocessed the data with the following methods: 
\begin{enumerate}
\item Discarded the segments with UNKNOWN and MOVEMENT labels.
\item Combined N4 and N3 together as N3 stage. 
\item Ignored wake epochs longer than 30 minutes outside of sleep periods.   
\end{enumerate}

\begin{table}[!ht]

    \centering
    \caption{data distribution}
    \renewcommand{\arraystretch}{1.2} 
        \begin{tabular}{|c|c|c|c|c|c|}
        \hline
                           & \multicolumn{1}{c|}{Wake}    & \multicolumn{1}{c|}{N1}      & \multicolumn{1}{c|}{N2}      & \multicolumn{1}{c|}{N3}  & REM     \\ \hline\linebreak
        Segment & \multicolumn{1}{c|}{44752}   & \multicolumn{1}{c|}{15793}   & \multicolumn{1}{c|}{54682}   & \multicolumn{1}{c|}{12268}  & 20976   \\ \hline\linebreak
        Distribution       & \multicolumn{1}{c|}{30.14\%} & \multicolumn{1}{c|}{10.64\%} & \multicolumn{1}{c|}{36.83\%} & \multicolumn{1}{c|}{8.26\%} & 14.13\% \\ \hline\linebreak
        Total Number       & \multicolumn{5}{c|}{148471}                                                                                                        \\ \hline
        \end{tabular}
    \label{table:data_distribution}
\end{table}

\subsection{Performance Metrics}
\par We evaluated our model's performance using per-class Precision (PR), per-class Recall (RE), per-class F1-score (F1), and overall accuracy (acc). Overall accuracy is the ratio between the number of correct predictions and the population. For a category prediction, there are four outcomes: true positive (TP), false positive (FP), true negative (TN), false negative (FP). Metrics are defined as:  
\begin{equation}
 PR = \frac{TP}{TP + FP}
\end{equation}
\begin{equation}
    RE = \frac{TP}{TP + FN}
\end{equation}
\begin{equation}
    F1 = \frac{2TP}{2TP + FN + FP}
\end{equation}
\par Overall accuracy is commonly used to measure classification performance. However, for an imbalanced dataset, precision does not provide adequate information on classifiers, because it hardly reveals performance in minority groups \cite{6}. Table ~\ref{table:data_distribution} shows the distribution of the dataset. The dataset is highly imbalanced, so we introduced additional metrics, PR, RE, and F1, to correctly measure the performance of our model.

\subsection{Proposed Model}
\par Raw EEG data contains time-invariant and time-variant features. Each 30-second input data segment with 100-Hz sampling frequency, so the input shape is (3000,1). It is too large to feed it into a transformer unit. Fig.~\ref{fig:model} shows our model's architecture. A convolutional neural network can extract time-invariant data and output smaller data. We implemented four sequential convolutional layers to output features with shape (19,128). Then we use a transformer unit to learn some time-variant information from the features. Its attention mechanism learned the contexts on all positions of the time series data. The two dense layers inside the transformer unit work as an encoder. The output of the encoder is then added to the input data for additional features. Finally, to correctly classify sleep scores, we used a dense layer with a softmax activation function to obtain the most possible categories. We also tried other models, such as recurrent neural network and auto-encoders. Experiments showed that our proposed model yielded the best performance.

\section{Experiments}
\subsection{Data Preprocessing}
\par Since our model was designed to be deployed on microcontrollers with limited memory and computation resources, we cannot design a complex data preprocessing method, so we implemented a simple standarization. The performance of deep-learning models is highly dependent on the statistical properties of the input data. If the input data is too small or too large, the models perform poorly. The EEG data in the Fpz-Cz channel is within the scope of $10^{-5}$, which is too small for our model. Standarization transforms data to have a zero mean and a standard deviation of one. For each sample $X$, we standardized it to $Z$ using \[Z = \frac{X-M}{S}\] 
where M indicates the mean of the samples and S indicates the standard deviation of the samples. After standarization, the input data was in the range of $10^{-1}$ to $10^1$, and the model showed the best performance. 

\subsection{Basic Training}
\par We used 5-fold cross-validation to train and test our model. There are 77 subjects in total, and each fold contains 16 subjects' data (one fold has 13 subjects). In each iteration, we selected four folds as training and validating data, and one fold as testing data. Among the four folds of data, we randomly sampled 10\% for validation. The remaining 90\% of data was for training. We used Adam algorithm for optimization, which is a stochastic gradient descent method based on adaptive estimation of first-order and second-order moments. It computes efficiently and requires little memory \cite{7}. We utilized categorical cross-entropy as the loss function and fed our model with a batch size of 64 samples. 

\subsection{Subject-Specific Training}
\par We also performed subject-specific training in the testing stage to further adapt the patterns of each subject. We randomly selected 10\% of the test data and fed them to our trained model. The remaining 90\% of the test data were used to test our models after subject-specific training. 

\section{Results}
\par Table \ref{table:conf_before} and Table \ref{table:pre_trained} show the confusion table and the performance of our model, respectively. Since the dataset is imbalanced, the per-class performance in N1 and REM was expected to be poorer than the majority classes. The model performed well in the Wake and N2 classes. The state-of-the-art performances from literature are also similar. From Table \ref{table:compare}, all models have an F1 score per class less than 0.5 in N1 and less than 0.8 in REM. To improve the performance of our model, we perform a subject-specific training to further adapt subject-wise patterns. Table \ref{table:conf_after} and Table \ref{table:after_trained} show that performances improved after subject-specific training. Firstly, the overall accuracy increased from 0.775 to 0.795. Secondly, per-class precision, recall, and F1 score increased for all classes.

\par Table \ref{table:compare} compares the F1 score of different models from the literature. Since our proposed model is lightweight and small, it cannot yield the best performance. However, it still had performance comparable to that of the state-of-the-art. For the wake stage, we yielded the highest F1-score of 0.91. For other classes, we were close to the highest. There is no model that could beat our performances in all classes. Some models performed better on certain classes.

\begin{table}[!ht]
\centering
\label{table:conf_before}
\renewcommand{\arraystretch}{1.2} 
	\setlength{\tabcolsep}{5pt} 
\caption{Confusion matrix before subject-specific training}
\begin{tabular}{|c|c|c|c|c|c|}
\hline
 {Actual/Predict} &
   {Wake} &
   {N1} &
   {N2} &
   {N3} &
   {REM} \\ \hline
{Wake} &
  \cellcolor[HTML]{00009B}{\color[HTML]{FFFFFF}  {0.90}} &
  \cellcolor[HTML]{FFFFFF} {0.04} &
  \cellcolor[HTML]{FFFFFF} {0.01} &
  \cellcolor[HTML]{FFFFFF} {0} &
  \cellcolor[HTML]{FFFFFF} {0.04} \\ \hline
 {N1} &
  \cellcolor[HTML]{C3D9FD} {0.20} &
  \cellcolor[HTML]{C3D9FD}{\color[HTML]{333333}  {0.26}} &
  \cellcolor[HTML]{6EA4FB} {0.31} &
  \cellcolor[HTML]{FFFFFF} {0.01} &
  \cellcolor[HTML]{C3D9FD} {0.21} \\ \hline
 {N2} &
  \cellcolor[HTML]{FFFFFF} {0.01} &
  \multicolumn{1}{l|}{\cellcolor[HTML]{FFFFFF} {0.04}} &
  \multicolumn{1}{l|}{\cellcolor[HTML]{3166FF}{\color[HTML]{FFFFFF}  {0.82}}} &
  \cellcolor[HTML]{FFFFFF} {0.07} &
  \multicolumn{1}{l|}{\cellcolor[HTML]{FFFFFF} {0.06}} \\ \hline
 {N3} &
  \cellcolor[HTML]{FFFFFF} {0} &
  \cellcolor[HTML]{FFFFFF} {0} &
  \cellcolor[HTML]{C3D9FD} {0.19} &
  \cellcolor[HTML]{3166FF}{\color[HTML]{FFFFFF}  {0.80}} &
  \cellcolor[HTML]{FFFFFF} {0} \\ \hline
 {REM} &
  \cellcolor[HTML]{FFFFFF} {0.05} &
  \cellcolor[HTML]{FFFFFF} {0.08} &
  \cellcolor[HTML]{FFFFFF} {0.12} &
  \cellcolor[HTML]{FFFFFF} {0} &
  \cellcolor[HTML]{53ACFD}{\color[HTML]{FFFFFF}  {0.74}} \\ \hline
\end{tabular}
\end{table}

\begin{table}[h!]
\centering
\caption{Performance before subject-specific training}
\renewcommand{\arraystretch}{1.2} 
	\setlength{\tabcolsep}{9pt} 
\begin{tabular}{|c|c|c|c|c|c|}
\hline\linebreak
                               & \multicolumn{1}{c|}{Wake} & \multicolumn{1}{c|}{N1}   & \multicolumn{1}{c|}{N2}   & \multicolumn{1}{c|}{N3} & REM                       \\ \hline
Precision                      & \multicolumn{1}{c|}{0.90} & \multicolumn{1}{c|}{0.26} & \multicolumn{1}{c|}{0.72} & \multicolumn{1}{c|}{0.80}  & 0.74                      \\ \hline
Recall                         & \multicolumn{1}{c|}{0.89} & \multicolumn{1}{c|}{0.42} & \multicolumn{1}{c|}{0.81} & \multicolumn{1}{c|}{0.70}  & 0.64                      \\ \hline
F1-Score                       & \multicolumn{1}{c|}{0.90} & \multicolumn{1}{l|}{0.32} & \multicolumn{1}{l|}{0.82} & \multicolumn{1}{l|}{0.75}  & \multicolumn{1}{l|}{0.69} \\ \hline
\multicolumn{1}{|l|}{Accuracy} & \multicolumn{5}{c|}{0.775}                                                                                                                 \\ \hline
\end{tabular}
\linebreak
\label{table:pre_trained}
\end{table}

\begin{table}[!ht]
\centering
\caption{Confusion matrix after subject-specific training}
\label{table:conf_after}
\renewcommand{\arraystretch}{1.2} 
	\setlength{\tabcolsep}{5pt} 
\begin{tabular}{|c|c|c|c|c|c|}
\hline
 {Actual/Predict} &
   {Wake} &
   {N1} &
   {N2} &
   {N3} &
   {REM} \\ \hline
 {Wake} &
  \cellcolor[HTML]{00009B}{\color[HTML]{FFFFFF}  {0.91}} &
  \cellcolor[HTML]{FFFFFF} {0.05} &
  \cellcolor[HTML]{FFFFFF} {0.01} &
  \cellcolor[HTML]{FFFFFF} {0} &
  \cellcolor[HTML]{FFFFFF} {0.02} \\ \hline
 {N1} &
  \cellcolor[HTML]{C3D9FD} {0.21} &
  \cellcolor[HTML]{6EA4FB}{\color[HTML]{FFFFFF}  {0.31}} &
  \cellcolor[HTML]{C3D9FD} {0.29} &
  \cellcolor[HTML]{FFFFFF} {0.01} &
  \cellcolor[HTML]{C3D9FD} {0.18} \\ \hline
 {N2} &
  \cellcolor[HTML]{FFFFFF} {0.01} &
  \multicolumn{1}{l|}{\cellcolor[HTML]{FFFFFF} {0.04}} &
  \multicolumn{1}{l|}{\cellcolor[HTML]{3166FF}{\color[HTML]{FFFFFF}  {0.84}}} &
  \cellcolor[HTML]{FFFFFF} {0.06} &
  \multicolumn{1}{l|}{\cellcolor[HTML]{FFFFFF} {0.05}} \\ \hline
 {N3} &
  \cellcolor[HTML]{FFFFFF} {0} &
  \cellcolor[HTML]{FFFFFF} {0} &
  \cellcolor[HTML]{C3D9FD} {0.20} &
  \cellcolor[HTML]{3166FF}{\color[HTML]{FFFFFF}  {0.80}} &
  \cellcolor[HTML]{FFFFFF} {0} \\ \hline
 {REM} &
  \cellcolor[HTML]{FFFFFF} {0.05} &
  \cellcolor[HTML]{FFFFFF} {0.08} &
  \cellcolor[HTML]{FFFFFF} {0.10} &
  \cellcolor[HTML]{FFFFFF} {0} &
  \cellcolor[HTML]{53ACFD}{\color[HTML]{FFFFFF}  {0.78}} \\ \hline
\end{tabular}
\end{table}

\begin{table}[h!]
\centering
\renewcommand{\arraystretch}{1.2} 
\caption{Performance after patient-specific training}
\begin{tabular}{|c|c|c|c|c|c|}
\hline\linebreak
                               & \multicolumn{1}{c|}{Wake} & \multicolumn{1}{c|}{N1}   & \multicolumn{1}{c|}{N2}   & \multicolumn{1}{c|}{N3} & REM                       \\ \hline\linebreak
Precision                      & \multicolumn{1}{c|}{0.92} & \multicolumn{1}{c|}{0.31} & \multicolumn{1}{c|}{0.84} & \multicolumn{1}{c|}{0.80}  & 0.78                      \\ \hline\linebreak
Recall                         & \multicolumn{1}{c|}{0.90} & \multicolumn{1}{c|}{0.46} & \multicolumn{1}{c|}{0.83} & \multicolumn{1}{c|}{0.74}  & 0.69                      \\ \hline\linebreak
F1-Score                       & \multicolumn{1}{c|}{0.91} & \multicolumn{1}{l|}{0.37} & \multicolumn{1}{l|}{0.84} & \multicolumn{1}{c|}{0.77}  & \multicolumn{1}{l|}{0.73} \\ \hline
\multicolumn{1}{|l|}{Accuracy} & \multicolumn{5}{c|}{0.795}                                                                                                                 \\ \hline
\end{tabular}
\linebreak

\label{table:after_trained}
\end{table}

\begin{table*}[!ht]
\centering
\caption{F1-Score comparison with state-of-art}
\renewcommand{\arraystretch}{1.2} 
\label{table:compare}
	\setlength{\tabcolsep}{9pt} 
\scalebox{1}{
\begin{tabular}{|ccc|ccccc|c|}
\hline
\multicolumn{1}{|c|}{\multirow{2}{*}{Reference}} & \multicolumn{1}{c|}{\multirow{2}{*}{Year}} & \multirow{2}{*}{Architecture} & \multicolumn{5}{c|}{F1-Score}                                                                                        & \multirow{2}{*}{Accuracy} \\ \cline{4-8}
\multicolumn{1}{|c|}{}                           & \multicolumn{1}{c|}{}                      &                               & \multicolumn{1}{c|}{Wake} & \multicolumn{1}{c|}{N1}   & \multicolumn{1}{c|}{N2}   & \multicolumn{1}{c|}{N3}   & REM  &                           \\ \hline
\multicolumn{1}{|c|}{\cite{9}}                        & \multicolumn{1}{c|}{2022}                  & DSSNet                        & \multicolumn{1}{c|}{0.86} & \multicolumn{1}{c|}{0.20} & \multicolumn{1}{c|}{0.87} & \multicolumn{1}{c|}{0.87} & 0.71 & 0.82                      \\ \hline
\multicolumn{1}{|c|}{\cite{10}}                       & \multicolumn{1}{c|}{2018}                  & 1-Max-CNN                     & \multicolumn{1}{c|}{0.77} & \multicolumn{1}{c|}{0.33} & \multicolumn{1}{c|}{0.87} & \multicolumn{1}{c|}{0.86} & 0.76 & 0.80                      \\ \hline
\multicolumn{1}{|c|}{\cite{3}}                        & \multicolumn{1}{c|}{2017}                  & DeepSleepNet                  & \multicolumn{1}{c|}{0.88} & \multicolumn{1}{c|}{0.37} & \multicolumn{1}{c|}{0.83} & \multicolumn{1}{c|}{0.77} & 0.80 & 0.80                      \\ \hline
\multicolumn{1}{|c|}{\cite{12}}                       & \multicolumn{1}{c|}{2019}                  & SleepEEGNet                   & \multicolumn{1}{c|}{0.91} & \multicolumn{1}{c|}{0.44} & \multicolumn{1}{c|}{0.82} & \multicolumn{1}{c|}{0.73} & 0.76 & 0.80                      \\ \hline
\multicolumn{1}{|c|}{\cite{13}}                       & \multicolumn{1}{c|}{2021}                  & CNN                           & \multicolumn{1}{c|}{0.91} & \multicolumn{1}{c|}{0.42} & \multicolumn{1}{c|}{0.77} & \multicolumn{1}{c|}{0.66} & 0.69 & N/A                       \\ \hline
\multicolumn{3}{|c|}{\textbf{This work}}                                                                                   & \multicolumn{1}{c|}{0.91} & \multicolumn{1}{c|}{0.37} & \multicolumn{1}{c|}{0.84} & \multicolumn{1}{c|}{0.77} & 0.73 & 0.80                      \\ \hline
\end{tabular}
}
\end{table*}

\section{Discussion}

\par Our lightweight model was designed for memory-constrained microcontroller units. It had around 300,000 parameters, and its size was 2 MB. We used an Arduino Nano 33 BLE board, which integrates a wireless microntroller (nRF52840, Nordic Semiconductor) with a 64 MHz 32-bit ARM CPU, 1 MB of flash memory, and 256 KB of SRAM \cite{arduino}. The deployed model was stored in the flash memory, so the model was supposed to be less than 1 MB. We planned to add an external memory unit to the model in the future. To test the availability of our design in the current stage, we implemented a smaller version of the model and deployed it on the Arduino board. The model had an accuracy of 68\%, but the microcontroller was fully functional for target classification. To reduce the size of our model, we also quantized the model in the deployment stage. 

We randomly selected 10\% of the test data set to perform subject-specific training. If a dataset is large, the randomly selected data should follow the distribution of the dataset. In our experiments, the distribution varied as the dataset was not large enough. The improvement of performance was highly dependent on the distribution of the selected data. Therefore, we will enforce the distribution of the subject-specific training data. This means that the number of selected data in each category is calculated and fixed based on the test dataset. This method would maximize the benefits of subject-specific training.

As mentioned previously, the dataset has an imbalanced class distribution, which significantly affects models' performance, especially on the minority categories. There are different methods to mitigate the effect of imbalance data, such as oversampling and undersampling. Oversampling is used to duplicate samples in minority classes, while undersampling is used to remove samples from the majority classes. There are two common ways in training imbalanced data. Another method we will try in the future is to add weights to the loss function. The loss function can be weighted differently for different classes, so that minority classes are learned more.

\section{Conclusion}
In this work, we developed a lightweight DL model for real-time sleep stage classification using single-channel EEG data. The DL model features CNN and transformers. We validated the model using the Sleep-EDF dataset and tested it in a low-power microcontroller. The model achieved performance comparable to the state-of-the-art works. In the future, we plan to develop a fully integrated wireless EEG sensor using the model. The developed device can enable a wide range of sleep research. 

\bibliographystyle{IEEEtran}
\bibliography{ref}

\end{document}